\documentclass[12pt,english]{article}
\usepackage{times}
\usepackage[T1]{fontenc}
\usepackage{geometry}
\usepackage{rotating}
\usepackage{graphicx}
\usepackage{setspace}
\usepackage{amssymb}

\makeatletter

\providecommand{\tabularnewline}{\\}

\usepackage{bm}

\usepackage{babel}
\makeatother
\begin{document}

\section*{\noindent Synthesis of Wide-Angle Scanning Arrays through Array Power
Control}

\noindent \vfill

\noindent P. Rosatti,$^{(1)(2)}$ G. Oliveri, $^{(1)(2)}$ \emph{Fellow,
IEEE}, and A. Massa,$^{(1)(2)(3)(4)(5)}$ \emph{Fellow, IEEE}

\noindent \vfill

\noindent {\footnotesize $^{(1)}$} \emph{\footnotesize ELEDIA Research
Center} {\footnotesize (}\emph{\footnotesize ELEDIA}{\footnotesize @}\emph{\footnotesize UniTN}
{\footnotesize - University of Trento)}{\footnotesize \par}

\noindent {\footnotesize DICAM - Department of Civil, Environmental,
and Mechanical Engineering}{\footnotesize \par}

\noindent {\footnotesize Via Mesiano 77, 38123 Trento - Italy}{\footnotesize \par}

\noindent \textit{\emph{\footnotesize E-mail:}} {\footnotesize \{}\emph{\footnotesize pietro.rosatti,
giacomo.oliveri, andrea.massa}{\footnotesize \}@}\emph{\footnotesize unitn.it}{\footnotesize \par}

\noindent {\footnotesize Website:} \emph{\footnotesize www.eledia.org/eledia-unitn}{\footnotesize \par}

\noindent {\footnotesize $^{(2)}$} \emph{\footnotesize CNIT - \char`\"{}University
of Trento\char`\"{} ELEDIA Research Unit }{\footnotesize \par}

\noindent {\footnotesize Via Sommarive 9, 38123 Trento - Italy}{\footnotesize \par}

\noindent {\footnotesize Website:} \emph{\footnotesize www.eledia.org/eledia-unitn}{\footnotesize \par}

\noindent {\footnotesize $^{(3)}$} \emph{\footnotesize ELEDIA Research
Center} {\footnotesize (}\emph{\footnotesize ELEDIA}{\footnotesize @}\emph{\footnotesize UESTC}
{\footnotesize - UESTC)}{\footnotesize \par}

\noindent {\footnotesize School of Electronic Science and Engineering,
Chengdu 611731 - China}{\footnotesize \par}

\noindent \textit{\emph{\footnotesize E-mail:}} \emph{\footnotesize andrea.massa@uestc.edu.cn}{\footnotesize \par}

\noindent {\footnotesize Website:} \emph{\footnotesize www.eledia.org/eledia}{\footnotesize -}\emph{\footnotesize uestc}{\footnotesize \par}

\noindent {\footnotesize $^{(4)}$} \emph{\footnotesize ELEDIA Research
Center} {\footnotesize (}\emph{\footnotesize ELEDIA@TSINGHUA} {\footnotesize -
Tsinghua University)}{\footnotesize \par}

\noindent {\footnotesize 30 Shuangqing Rd, 100084 Haidian, Beijing
- China}{\footnotesize \par}

\noindent {\footnotesize E-mail:} \emph{\footnotesize andrea.massa@tsinghua.edu.cn}{\footnotesize \par}

\noindent {\footnotesize Website:} \emph{\footnotesize www.eledia.org/eledia-tsinghua}{\footnotesize \par}

\noindent {\small $^{(5)}$} {\footnotesize School of Electrical Engineering}{\footnotesize \par}

\noindent {\footnotesize Tel Aviv University, Tel Aviv 69978 - Israel}{\footnotesize \par}

\noindent \textit{\emph{\footnotesize E-mail:}} \emph{\footnotesize andrea.massa@eng.tau.ac.il}{\footnotesize \par}

\noindent {\footnotesize Website:} \emph{\footnotesize https://engineering.tau.ac.il/}{\footnotesize \par}

\noindent \vfill

\noindent {\small ~}{\small \par}

\noindent \emph{\small This work has been submitted to the IEEE for
possible publication. Copyright may be transferred without notice,
after which this version may no longer be accessible.}{\small \par}

\noindent \vfill

\newpage
\section*{Synthesis of Wide-Angle Scanning Arrays through Array Power Control}

~

~

~

\begin{flushleft}P. Rosatti, G. Oliveri, and A. Massa\end{flushleft}

\vfill

\begin{abstract}
\noindent A new methodology for the synthesis of wide-angle scanning
arrays is proposed. It is based on the formulation of the array design
problem as a multi-objective one where, for each scan angle, both
the radiated power density in the scan direction and the total reflected
power are accounted for. A set of numerical results from full-wave
simulated examples - dealing with different radiators, arrangements,
frequencies, and number of elements - is reported to show the features
of the proposed approach as well as to assess its potentialities.

\vfill
\end{abstract}
\noindent \textbf{Key words}: Wide-Angle Scanning; Excitation Design;
Array Synthesis; Active Reflection Coefficient Stabilization; Multi-Objective
Evolutionary Optimization.

\newpage
\section{Introduction and Rationale\label{sec:Introduction}}

\noindent Wide-angle scanning is one of the key requirements in modern
antenna arrays for communications and sensing \cite{Herd 2016}\cite{Gao 2022}.
Indeed, stable radiation performance over a wide range of steering
angles are increasingly important in several applicative scenarios
such as wireless coverage, automotive radars, satellite communications,
and imagers \cite{Herd 2016}-\cite{Oliveri 2022}. Unfortunately,
the gain/directivity of antenna arrays typically worsens as the scan
angle increases, specifically in the presence of {}``blind spots''
\cite{Herd 2016}\cite{Gao 2022}, and it limits the array field-of-view
(\emph{FoV}) \cite{Herd 2016}\cite{Gao 2022}. These undesired phenomena
are a direct consequence of the instability of the active return loss
(\emph{ARL}) versus the scan angle \cite{Hansen.2009}. 

\noindent In the last decades, several approaches have been introduced
in the state-of-the-art (\emph{SoA}) literature to improve the \emph{ARL}
stability and to increase the \emph{FoV} of phased arrays by mitigating
the inter-element coupling \cite{Oliveri 2022}\cite{Valavan 2014}-\cite{Oliveri 2021}.
More in detail, the exploitation of ad-hoc solutions such as the use
of cavities \cite{Valavan 2014}\cite{Valavan 2014b} or magnetic
dipoles \cite{Liu.2018} has been proposed to individually insulate
the array elements. An alternative strategy, not involving complex
radiators, resorts to the introduction of decoupling feed networks
\cite{Xia.2014}\cite{Xia.2015}\cite{Hannan.1965}, which synthesis
and implementation become very challenging when a wideband working
is required \cite{Xia.2014}\cite{Xia.2015}\cite{Hannan.1965}. Wide-angle
impedance matching (\emph{WAIM}) layers coating the array and based
either on standard \cite{Magill.1966} or artificial materials \cite{Oliveri 2022}\cite{Sajuyugbe 2010}-\cite{Oliveri 2021}
has been also adopted to stabilize the \emph{ARL}. Nonetheless, this
latter solution - as well as the previous ones - needs non-trivial
updates of the system hardware, as applied to existing architectures,
and customized designs/implementations strongly depend on the adopted
antenna technology \cite{Oliveri 2022}\cite{Sajuyugbe 2010}-\cite{Oliveri 2021}.

\noindent Otherwise, the \emph{ARL} stabilization may be yielded without
architectural modifications by considering a completely different
perspective, as well. Towards this end, let us notice that the \emph{ARL}
depends not only on the inter-element coupling, but also on the excitations
applied to each elementary radiator of the array \cite{Hansen.2009}\cite{Pozar.1994}.
While the synthesis of such excitations is typically aimed at controlling
the scan direction, the sidelobe profile \cite{Mailloux 2005}, and/or
more advanced quality-of-service indicators \cite{Oliveri.2019}\cite{Oliveri.2022b},
there are not theoretical obstacles to even address the stabilization
the \emph{ARL} still fulfilling the user-desired pattern features
when scanning.

\noindent Following this line of reasoning, an innovative methodology
for the synthesis of wide-scanning arrays is proposed into the following.
Such an approach is based on the formulation of the array synthesis
as a multi-objective problem where, for each scan angle, both the
radiated power density in the scan direction and the total reflected
power are accounted for by including the finite array coupling in
the design process (i.e., no periodicity approximation is employed).

\noindent Accordingly, the main innovative contributions of this work
include (\emph{i}) for the first time to the best of the authors'
knowledge, the synthesis of the array excitations to minimize the
reflected power, while keeping suitable radiation performance, for
widening the \emph{FoV} in finite arrays, (\emph{ii}) the full-wave
numerical proof for both linear and planar arrays that it is possible
to obtain wider scan angles without modifying the antenna architecture,
hence implicitly the assessment of the applicability of the proposed
concept to upgrade existing arrays, and (\emph{iii}) the derivation
of a general-purpose strategy for wide angle scanning that, unlike
\emph{SoA} methods \cite{Oliveri 2022}\cite{Valavan 2014}-\cite{Oliveri 2021},
can be adopted regardless of the underlying antenna technology and/or
physical implementation details.

\noindent The outline of the paper is as follows. The mathematical
formulation of the array synthesis problem at hand is formulated in
Sect. \ref{sec:Problem-Formulation}, where the proposed excitation-based
approach to yield wide-angle scanning performance is presented, as
well. A set of numerical results from full-wave simulated examples
is reported to show the features of the proposed strategy as well
as to assess its potentialities in different working conditions and
also in comparison with state-of-the-art competitive techniques (Sect.
\ref{sec:Results}). Finally, some conclusions are drawn (Sect. \ref{sec:Conclusions-and-Remarks}).

\section{\noindent Mathematical Formulation\label{sec:Problem-Formulation} }

\noindent Let us consider the scenario in Fig. 1 where an antenna
array of $N$ elements, each one described by the $L$ geometrical
parameters $\underline{g}=\left\{ g_{l};\, l=1,...,L\right\} $ and
operating at the carrier frequency $f_{0}$, is fed by the set of
$N$ incident excitation coefficients $\underline{w}_{q}^{+}\triangleq\left\{ w_{nq}^{+};\, n=1,...,N\right\} $
to radiate a beampattern steered at the $q$-th ($q=1,...,Q$) scan
angle $\left(\theta_{q},\varphi_{q}\right)$ \cite{Hansen.2009}\cite{Pozar.1994}\cite{Pozar.2003}.
The corresponding reflected excitation coefficients, measured at the
$n$-th ($n=1,...,N$) element, of the array are defined as \cite{Pozar.1994}\begin{equation}
w_{nq}^{-}\triangleq w_{nq}^{+}\times\Gamma_{nq}\label{eq:reflected excitation}\end{equation}
where\begin{equation}
\Gamma_{nq}\triangleq\sum_{m=1}^{N}S_{nm}\frac{w_{mq}^{+}}{w_{nq}^{+}}\label{eq:active RL}\end{equation}
it the $n$-th ($n=1,...,N$) element active reflection coefficient
at the $q$-th ($q=1,...,Q$) scan angle, the corresponding \emph{ARL}
being $ARL_{nq}\triangleq\left|\Gamma_{nq}\right|^{2}$ ($n=1,...,N$;
$q=1,...,Q$) \cite{Oliveri 2022}\cite{Hansen.2009}\cite{Sajuyugbe 2010}-\cite{Oliveri 2021},
and $S_{nm}$ is the ($n$, $m$)-th ($n$, $m=1,...,N$) entry of
the $N\times N$ scattering matrix $\mathcal{S}$ \cite{Hansen.2009}\cite{Pozar.1994}\cite{Mailloux 2005}\cite{Pozar.2003}.
This latter univocally models the array inter-element coupling effects
that, unlike $\Gamma_{nq}$ (\ref{eq:active RL}) and $w_{nq}^{-}$
(\ref{eq:reflected excitation}), does not depend on the applied array
excitations $w_{nq}^{+}$.

\noindent Under the above assumptions, the total excitation at the
$n$-th ($n=1,...,N$) array element, $w_{nq}$ ($w_{nq}\triangleq w_{nq}^{+}+w_{nq}^{-}$)
turns out to be \cite{Pozar.1994}\cite{Pozar.2003}\begin{equation}
w_{nq}=\left(1+\Gamma_{nq}\right)w_{nq}^{+},\label{eq:total excitations}\end{equation}
then field radiated radiated by the $N$-elements antenna array in
the Fraunhofer region has the following expression \cite{Pozar.1994}\cite{Mailloux 2005}\cite{Pozar.2003}\begin{equation}
\mathbf{E}\left(\left.\mathbf{r}\right|\underline{w}_{q}\right)=\frac{jk_{0}}{2\pi}\frac{\exp\left(-jk_{0}r\right)}{r}\sum_{n=1}^{N}w_{nq}\exp\left[-jk_{0}\left(\mathbf{r}_{n}\cdot\widehat{\mathbf{r}}\right)\right]\mathbf{E}_{n}\left(\theta,\varphi\right),\label{eq:radiated field}\end{equation}
which holds true regardless of the array geometry (linear/planar)
and element polarization (i.e., single/dual polarization). In (\ref{eq:radiated field}),
$k_{0}$ is the free-space wavenumber, $\mathbf{r}=\left\{ r,\theta,\varphi\right\} $
is the spatial coordinate ($\widehat{\mathbf{r}}=\frac{\mathbf{r}}{\left|\mathbf{r}\right|}$),
$\cdot$ stands for the scalar product, and $\mathbf{r}_{n}$ is the
location of the $n$-th ($n=1,...,N$) array element, which element
pattern is $\mathbf{E}_{n}\left(\theta,\varphi\right)$ \cite{Mailloux 2005}.
It is worth pointing out that in finite arrays, the element pattern
is different for each $n$-th ($n=1,...,N$) element {[}i.e., $\mathbf{E}_{n}\left(\theta,\varphi\right)\ne\mathbf{E}_{m}\left(\theta,\varphi\right)$,
$n\ne m$ ($n$, $m=1,...,N$){]}, even though the centrally-located
radiators of large arrays exhibit very similar angular responses \cite{Mailloux 2005}.

\noindent The power density radiated along the direction $\left(\theta,\varphi\right)$
by the array steered towards $\left(\theta_{q},\varphi_{q}\right)$
is obtained from (\ref{eq:radiated field}) as follows\begin{equation}
\psi_{q}\left(\theta,\varphi\right)=\lim_{r\to\infty}\frac{r^{2}}{2\nu}\left|\mathbf{E}\left(\left.\mathbf{r}\right|\underline{w}_{q}\right)\right|^{2}\label{eq:}\end{equation}
and the total radiated power is $\mathbb{P}_{q}^{RAD}=\int_{\Omega}\psi_{q}\left(\theta,\varphi\right)d\Omega$,
$\Omega$ being a far-field closed surface surrounding the array.
Moreover, the total reflected power percentage is defined as\begin{equation}
\zeta_{q}=\frac{\mathbb{P}_{q}^{REFL}}{\mathbb{P}_{q}^{IN}}\label{eq:}\end{equation}
where $\mathbb{P}_{q}^{IN}$ and $\mathbb{P}_{q}^{REFL}$ are the
input power ($\mathbb{P}_{q}^{IN}\triangleq\sum_{n=1}^{N}\left|w_{nq}^{+}\right|^{2}$)
and the reflected power ($\mathbb{P}_{q}^{REFL}\triangleq\sum_{n=1}^{N}\left|w_{nq}^{-}\right|^{2}$
equal to $\mathbb{P}_{q}^{REFL}\triangleq\sum_{n=1}^{N}ARL_{nq}\left|w_{nq}^{+}\right|^{2}$)
when the array is steered along $\left(\theta_{q},\varphi_{q}\right)$,
respectively. According to these definitions, the \emph{FoV} of an
antenna array can be identified as the angular range $\Omega_{FoV}$
($\Omega_{FoV}\subset\Omega_{TOT}$, $\Omega_{TOT}=\left\{ \left(\theta_{q},\varphi_{q}\right);\, q=1,...,Q\right\} $)
where the radiated power density in the scan direction $\left(\theta_{q},\varphi_{q}\right)$
is greater than a threshold $\psi_{TH}$ (i.e., $\psi_{q}\left(\theta_{q},\varphi_{q}\right)\ge\psi_{TH}$)
and the total reflected power percentage is smaller than a target
value $\zeta_{TH}$ (i.e., $\zeta_{q}\le\zeta_{TH}$). Mathematically,
it turns out that $\left(\theta_{q},\varphi_{q}\right)\in\Omega_{FoV}$
if\begin{equation}
\begin{array}{c}
\psi_{q}\left(\theta_{q},\varphi_{q}\right)\ge\psi_{TH}\\
\zeta_{q}\le\zeta_{TH}.\end{array}\label{eq:FoV}\end{equation}
In order to achieve wide scan performance (i.e., a wide \emph{FoV}),
\emph{SoA} methods in the reference literature \cite{Oliveri 2022}\cite{Valavan 2014}-\cite{Oliveri 2021}
require an $ARL_{nq}$ almost independent on the angular direction
(i.e., $ARL_{nq}\approx ARL_{n}$) and smaller than an application-defined
threshold $ARL_{TH}$ ($ARL_{n}\le ARL_{TH}$) by enforcing the condition
$\left|S_{nm}\right|\to0$ ($n=1,...,N$; $m=1,...,M$; $m\neq n$)
to yield\begin{equation}
\Gamma_{n}\left(\theta_{q},\varphi_{q}\right)\approx S_{nn}.\label{eq:}\end{equation}
Therefore, the condition $ARL_{n}\le ARL_{TH}$ is fulfilled by properly
setting the descriptors of the elementary radiator, $\underline{g}$,
so that $\left|S_{nn}\right|^{2}\le ARL_{TH}$, while the incident
excitation coefficients $\underline{w}_{q}^{+}$ are optimized to
fulfill the requirements on the radiated power density along $\left(\theta_{q},\varphi_{q}\right)$,
$\psi_{q}\left(\theta_{q},\varphi_{q}\right)$.

\noindent Otherwise, the proposed approach leverages the observation
that, according to (\ref{eq:active RL}), for a given matrix $\mathcal{S}$
(i.e., resulting from the definition of the element geometries, physical
properties, and positions), both the total reflected power percentage,
$\zeta_{q}$, and the radiation properties of the array $\mathbf{E}\left(\left.\mathbf{r}\right|\underline{w}_{q}\right)$
(\ref{eq:radiated field}) can be controlled by properly optimizing
the excitations $\underline{w}_{q}^{+}$ ($q=1,...,Q$) to maximize
the \emph{FoV} of the antenna array. Therefore, the design of the
optimal excitations for the $q$-th ($q=1,...,Q$) scan angle, $\left.\underline{w}_{q}^{+}\right\rfloor _{opt}$,
is mathematically formulated as a multi-objective constrained optimization
where $\left.\underline{w}_{q}^{+}\right\rfloor _{opt}$ is the set
of excitations $\underline{w}_{q}^{+}$ that fulfils some feasibility
conditions (i.e., $\underline{w}^{+}\in\mathcal{W}$, $\mathcal{W}$
being a set of requirements on the excitation coefficients such as
constant magnitude, phase quantization, etc ...), while minimizing
the $T=2$ objective functions coding the \emph{FoV} condition (\ref{eq:FoV})\begin{equation}
\left.\underline{w}_{q}^{+}\right\rfloor _{opt}\triangleq\arg\left\{ \min_{\underline{w}_{q}^{+}}\left[\Phi_{REFL}\left(\underline{w}_{q}^{+}\right);\,\Phi_{RAD}\left(\underline{w}_{q}^{+}\right)\right]\right\} .\label{eq:cost function generale}\end{equation}
In (\ref{eq:cost function generale}), the first objective function
$\Phi_{REFL}$ coincides with the total reflected power percentage
when the array is steered along $\left(\theta_{q},\varphi_{q}\right)$\begin{equation}
\Phi_{REFL}\left(\underline{w}_{q}^{+}\right)\triangleq\zeta_{q},\label{eq:Rsciugni}\end{equation}
while the second one, $\Phi_{RAD}$, is defined as the inverse of
the power density radiated along the $q$-th ($q=1,...,Q$) steering
direction\begin{equation}
\Phi_{RAD}\left(\underline{w}_{q}^{+}\right)=\frac{1}{\psi_{q}\left(\theta_{q},\varphi_{q}\right)}.\label{eq:Masciulla}\end{equation}
As it can be noticed, the synthesis process at hand does not involve
the geometrical descriptors of the array, $\underline{g}$, which
are user-defined and \emph{a-priori} set.

\noindent Since (\ref{eq:cost function generale}) codes a multi-objective
optimization problem, which involves highly non-linear cost function
terms, and the unknowns (i.e., magnitude/phase of the $\underline{w}^{+}$
entries) are real-valued, a multi-objective steady-state evolutionary
algorithm that uses $\varepsilon$-dominance archiving ($\varepsilon$\emph{-MOEA})
to record a diverse set of Pareto optimal solutions has been adopted
\cite{Deb.2003}-\cite{Rosatti 2023}.

\noindent Unlike generational algorithms \cite{Rocca 2009w}, the
$\varepsilon$\emph{-MOEA} evolves one solution at a time to ensure
convergence and diversity throughout the optimization process \cite{Deb.2003}-\cite{Rosatti 2023}.
For each $q$-th ($q=1,...,Q$) scan angle, $\left(\theta_{q},\varphi_{q}\right)$,
the following iterative procedure ($c$ being the iteration index)
is performed. At each $c$-th ($c=1,...,C$) iteration, both a variable-sized
Pareto front $\mathcal{A}_{cq}$ of $A_{cq}$ solutions {[}$\mathcal{A}_{cq}=\left\{ \left.\underline{w}_{q}^{+}\right\rfloor _{c}^{\left(a\right)};\, a=1,...,A_{cq}\right\} ${]}
and a fixed size population $\mathcal{P}_{cq}$ of $P$ individuals
{[}$\mathcal{P}_{cq}=\left\{ \left.\underline{w}_{q}^{+}\right\rfloor _{c}^{\left(p\right)};\, p=1,...,P\right\} ${]}
are evolved according to the $\varepsilon$\emph{-MOEA} mutation,
recombination, and selection operators with control parameters $\underline{\varepsilon}=\left\{ \varepsilon_{1},\varepsilon_{2}\right\} $
\cite{Deb.2003}-\cite{Rosatti 2023}. At the convergence (i.e., $c=C$),
a Pareto front of non-dominated excitation sets $\mathcal{A}_{Cq}$
is outputted and the user identifies the optimal one according to
a criterion based on either functional (i.e., a set of conditions
on $\Phi_{RAD}$ and/or $\Phi_{REFL}$) or non-functional (e.g., costs,
maintenance, etc ...) constraints.

\noindent Finally, the \emph{FoV} $\Omega_{FoV}$ of the synthesized
array (i.e., $\left.\underline{\underline{W}}^{+}\right\rfloor _{opt}=\left\{ \left.\underline{w}_{q}^{+}\right\rfloor _{opt};\, q=1,...,Q_{FoV}\right\} $)
turns out to be the set of $Q_{FoV}$ ($Q_{FoV}\le Q$) adjacent scan
directions for which (\ref{eq:FoV}) holds true.

\section{\noindent Numerical Results\label{sec:Results} }

\noindent This section is aimed at both illustrating the features
of the proposed method and assessing its effectiveness also in comparison
with state-of-the-art solutions.

\noindent The first numerical example deals with a linear array of
$N=5$ equally-spaced by $d=0.47\lambda$ cavity-backed slot-fed microstrip
antennas {[}Fig. 2(\emph{a}){]} printed on a Polyflon CuFlon substrate
with thickness $9\times10^{-3}$ {[}m{]}. Each radiator {[}Fig. 2(\emph{b}){]}
has been modeled in Ansys HFSS \cite{HFSS 2021} by choosing the following
$L=7$ descriptors to operate at $f=2$ GHz: $g_{1}=4.13\times10^{-2}$
{[}m{]}, $g_{2}=6.18\times10^{-2}$ {[}m{]}, $g_{3}=3.09\times10^{-2}$
{[}m{]}, $g_{4}=4.58\times10^{-2}$ {[}m{]}, $g_{5}=2.54\times10^{-3}$
{[}m{]}, $g_{6}=1.27\times10^{-2}$ {[}m{]}, and $g_{7}=4.63\times10^{-3}$
{[}m{]}. Moreover, $\Omega_{FoV}$ has been defined by setting in
(\ref{eq:FoV}) $\zeta_{TH}=10$ \% and choosing $\psi_{TH}$ so that
the maximum array scan loss is $-6$ {[}dB{]} (i.e., $\psi_{TH}=-6\,+\psi_{1}\left(90,\,0\right)$
{[}dB{]}). As for the array synthesis method, the requirement that
no magnitude tapering is allowed (i.e., $\left|w_{nq}^{+}\right|=1$,
$n=1,...,N$ and $q=1,...,Q$), so that only the phases of the excitations
are optimized (\emph{PO}), has been encoded into the feasibility set
$\mathcal{W}$ and the criterion for selecting the optimal trade-off
array (i.e., $\left.\underline{\underline{W}}^{+}\right\rfloor _{opt}$)
has been defined by setting $\Phi_{RAD}^{PO}\equiv\Phi_{RAD}^{STD}$,
$\Phi_{RAD}^{STD}$ being (\ref{eq:Masciulla}) of the standard approach
(\emph{STD}) where the excitation phases are computed with the analytic
linear phase shifting rule \cite{Mailloux 2005}. Finally, the $\varepsilon$\emph{-MOEA}
optimizer has been configured according to reference guidelines \cite{Deb.2003}-\cite{Rosatti 2023}
by choosing the following setup: $\left\{ \varepsilon_{1},\varepsilon_{2}\right\} =\left\{ 5\times10^{-3},2.5\times10^{-2}\right\} $,
$P=50$, and $C=1000$.

\noindent In order to illustrate the features and the behavior of
the proposed method for a generic $q$-th ($q=1,...,Q$) scan angle,
Figure 2(\emph{c}) shows the Pareto front $\mathcal{A}_{Cq}$ when
$\left(\theta_{q},\varphi_{q}\right)=\left(90,\,52\right)$ {[}deg{]}
($\to$ $q=53$). As it can be observed the \emph{PO} Pareto front
dominates the \emph{STD} solution and the optimal trade-off array
{[}i.e., the {}``green'' cross in Fig. 2(\emph{c}){]} reduces the
total reflected power percentage from $\zeta_{q}^{STD}=24.6$ \% (i.e.,
$\Phi_{REFL}^{STD}=0.246$) down to $\zeta_{q}^{PO}=8.9$ \% (i.e.,
$\Phi_{REFL}^{PO}=0.089$), while keeping the same radiation performance
(i.e., $\left.\psi_{q}^{PO}\left(\theta_{q},\varphi_{q}\right)\right\rfloor _{q=53}=\left.\psi_{q}^{STD}\left(\theta_{q},\varphi_{q}\right)\right\rfloor _{q=53}=34.6$
{[}dBW/sterad{]} $\to$ $\Phi_{RAD}=3.43\times10^{-4}$). The optimized
excitations, which are reported in Fig. 3(\emph{a}), turns out to
be very similar, which implies an analogous architectural complexity,
while the corresponding \emph{ARL} values are mitigated with respect
to the \emph{STD} approach {[}$ARL_{nq}^{PO}<ARL_{nq}^{STD}$, $n=1,...,N$
- Fig. 3(\emph{b}){]}. This latter outcome confirms that it is possible
to indirectly control the \emph{ARL} of a finite array by minimizing
the total reflected power without reducing the radiated power along
the steering direction as shown in Fig. 4.

\noindent The resume of the \emph{PO} performance for the first test
case is given in Fig. 5 where the plots of the radiated power density,
$\psi_{q}\left(\theta_{q},\varphi_{q}\right)$ ($q=1,...,Q$; $Q=91$)
{[}Fig. 5(\emph{a}){]}, and the total reflected power percentage,
$\zeta_{q}$ ($q=1,...,Q$; $Q=91$) {[}Fig. 5(\emph{b}){]}, are reported.
As it can be inferred, the scan width of the finite array is enlarged
of about $22$ {[}deg{]} (i.e., $-42$ {[}deg{]} $\le\Omega_{FoV}^{STD}\le$
$42$ {[}deg{]} vs. $-53$ {[}deg{]} $\le\Omega_{FoV}^{PO}\le$ $53$
{[}deg{]}), which corresponds to an increment of the field of view
percentage $\alpha_{FoV}$ ($\alpha_{FoV}\triangleq\frac{Q_{FoV}}{Q}$)
from $\alpha_{FoV}^{STD}\approx46$ \% up to $\alpha_{FoV}^{PO}\approx58$
\% ($\to$ $\Delta\alpha_{FoV}\approx12$ \%) and a maximum reduction
of the reflected power within $\Omega_{FoV}^{PO}$ that amounts to
$\Delta\zeta_{max}^{STD}=15.7$ \% {[}$\Delta\zeta_{max}^{STD}\triangleq\max_{q=1,...,Q_{FoV}}\left(\Delta\zeta_{q}^{STD}\right)$
being $\Delta\zeta_{q}^{STD}\triangleq\zeta_{q}^{STD}-\zeta_{q}^{PO}${]}.

\noindent The applicability of the considered approach to wider linear
arrays is assessed next by considering a larger arrangement with $N=16$
elements {[}Fig. 6(\emph{a}){]}. While there is still a widening of
the \emph{FoV} {[}Figs. 6(\emph{b})-6(\emph{c}){]}, the improvement
with respect to the \emph{STD} layout is here smaller ($\Delta\alpha_{FoV}\approx4$
\% $\to$ $\Delta\Omega_{FoV}\approx8$ {[}deg{]}) since the efficiency
of the proposed method in balancing the input power among the array
elements is vanified by the uniform coupling among the majority of
them. 

\noindent The third test case is concerned with planar architectures.
Towards this end, a planar $N=4\times4$ arrangement of radiators
as in Fig. 2(\emph{a}) has been considered (Fig. 7 - $d_{x}=d_{y}=0.47\,\lambda$)
and the results from the \emph{STD} and the \emph{PO} syntheses are
summarized in Fig. 8 by showing the color maps of $\psi_{q}\left(\theta_{q},\varphi_{q}\right)$
($q=1,...,Q$; $Q=361$) {[}Figs. 8(\emph{a})-8(\emph{b}){]} and $\zeta_{q}$
($q=1,...,Q$; $Q=361$) {[}Figs. 8(\emph{c})-8(\emph{d}){]}. As expected,
Figs. 8(\emph{a})-8(\emph{b}) are almost identical by definition since
the criterion for selecting the {}``optimal'' trade-off solution
forces the condition $\Phi_{RAD}^{PO}\approx\Phi_{RAD}^{STD}$. Otherwise,
the advantage of the \emph{PO} approach is evident when comparing
Fig. 8(\emph{c}) and Fig. 8(\emph{d}) and it is further highlighted
in Fig. 8(\emph{e}) where the reduction of the total reflected power,
$\Delta\zeta_{q}^{STD}$ ($q=1,...,Q$; $Q=361$) is reported, $\Delta\zeta_{max}$
being equal to $\Delta\zeta_{max}\approx8$ \%. Therefore, the main
outcome (i.e., the scanning performance of the finite array) is that
the \emph{PO} widens the field of view percentage from $\alpha_{FoV}^{STD}\approx40$
\% {[}Fig. 9(\emph{a}){]} up to $\alpha_{FoV}^{PO}\approx49$ \% {[}Fig.
9(\emph{b}){]} ($\to$ $\Delta\alpha_{FoV}\approx9$ \%).

\noindent The final numerical experiment is aimed at further assessing
the \emph{PO}-based solutions also against state-of-the-art wide scan
layouts \cite{Cheng.2017}. Accordingly, a planar arrangement of $N=8\times8$
($d_{x}=0.42\lambda$, $d_{y}=0.43\lambda$) identical parasitic pixel
layer-based E-shaped radiators \cite{Cheng.2017} operating at $f=5.2$GHz
has been numerically modeled in Ansys HFSS (Fig. 10). 

\noindent For comparison purposes, the plots of both the radiated
power density and the total reflected power along the cut $\varphi_{q}=0$
{[}deg{]} are shown in Fig. 11 where the values in Tab. I \cite{Cheng.2017}
are reported, as well. Once again the \emph{PO} array outperforms
the \emph{STD} one yielding on the $\varphi_{q}=0$ {[}deg{]} plane
an improvement of the \emph{FoV} of about $\Delta\alpha_{FoV}\approx13.2$
\% ($\alpha_{FoV}^{STD}\approx35.4$ \% vs. $\alpha_{FoV}^{PO}\approx48.6$
\% {[}Figs. 11(\emph{a})-11(\emph{b}){]}), while it turns out to be
$\Delta\alpha_{FoV}\approx22.6$ \% in the whole angular range {[}Fig.
12(\emph{a}) vs. Fig. 12(\emph{b}){]}. 

\noindent The \emph{PO} layout is also better than that in \cite{Cheng.2017}
in terms of $\psi_{q}\left(\theta_{q},0\right)$ and $\zeta_{q}$
($q=1,...,Q$; $Q=37$) at the available angular samples (i.e., $\theta_{4}=-75$
{[}deg{]}, $\theta_{8}=-55$ {[}deg{]}, $\theta_{30}=55$ {[}deg{]},
and $\theta_{34}=75$ {[}deg{]}) since $2.49$ {[}dB{]} $\le\Delta\psi_{q}^{[Cheng\,2017]}\left(\theta_{q},0\right)\le$
$18.35$ {[}dB{]} ($\Delta\psi_{q}^{[Cheng\,2017]}\left(\theta_{q},\varphi_{q}\right)\triangleq\psi_{q}^{PO}\left(\theta_{q},\varphi_{q}\right)-\psi_{q}^{[Cheng\,2017]}\left(\theta_{q},\varphi_{q}\right)$
{[}dB{]}) and $8.8$ \% $\le\Delta\zeta_{q}^{[Cheng\,2017]}\le$ $24.64$
\% in Tab. I. For fairness, the comparison has been also carried out
by considering the figures-of-merit optimized in \cite{Cheng.2017}
and the outcomes are summarized in Tab. II where $SLL_{q}$, $G_{q}$,
and $\Delta\Theta_{q}$ are the normalized sidelobe level ($SLL_{q}\triangleq\frac{\max_{\left(\theta,\varphi\right)\notin ML}\psi_{q}\left(\theta,\varphi\right)}{\max_{\left(\theta,\varphi\right)}\psi_{q}\left(\theta,\varphi\right)}$,
$ML$ being the main-lobe region), the array gain ($G_{q}\triangleq4\pi\frac{\max_{\left(\theta,\varphi\right)}\psi_{q}\left(\theta,\varphi\right)}{\mathbb{P}_{q}^{IN}}$),
and the scan angle error ($\Delta\Theta_{q}\triangleq\theta_{q}-\arg\left\{ \max_{\left(\theta,\varphi\right)}\psi_{q}\left(\theta,\varphi\right)\right\} $),
respectively. As it can be noticed (Tab. II), the values for the two
layouts are quite similar (i.e., $-3.66$ {[}dB{]} $\le\Delta SLL_{q}^{[Cheng\,2017]}\le$
$-0.23$ {[}dB{]}, $0.01$ {[}dB{]} $\le\Delta G_{q}^{[Cheng\,2017]}\le$
$0.50$ {[}dB{]}, and $-0.48$ {[}deg{]} $\le\Delta\Theta_{q}^{[Cheng\,2017]}\le$
$4.12$ {[}deg{]}).

\noindent Finally, Figure 13 gives the color level maps of the \emph{ARL}
of the $N=8\times8$ array elements when $\left(\theta_{q},\varphi_{q}\right)=\left(-75,0\right)$
{[}deg{]} (i.e., $q=4$), $ARL_{nq}$ ($n=1,...,N$), for the \emph{PO}
{[}Fig. 13(\emph{a}){]}, the \emph{SoA} \cite{Cheng.2017} {[}Fig.
13(\emph{b}){]}, and the \emph{STD} {[}Fig. 13(\emph{c}){]} layouts.
One can observe that generally $ARL_{nq}^{PO}<ARL_{TH}$ ($n=1,...,N$;
$q=4$), $ARL_{TH}=-10$ {[}dB{]}. This outcome further assesses the
capability of the \emph{PO}-based synthesis to indirectly control
the \emph{ARL} of a finite array by minimizing the total reflected
power without reducing the radiated power along the steering direction.
Moreover, the condition $ARL_{nq}^{PO}<ARL_{nq}^{[Cheng\,2017]}$
($q=4$) holds true in several $n$-th ($n=1,...,N$) radiating elements
{[}Fig. 13(\emph{a}) vs. Fig. 13(\emph{b}){]}, while $ARL_{nq}^{[Cheng\,2017]}\approx ARL_{nq}^{STD}$
($n=1,...,N$; $q=4$).

\section{\noindent Conclusions and Observations\label{sec:Conclusions-and-Remarks}}

\noindent The synthesis of the excitations of antenna arrays to achieve
wide-angle scanning features {[}i.e., a wide field-of-view (\emph{FoV}){]}
with minimum reflected power has been addressed. Towards this end,
an innovative methodology has been proposed by formulating the array
synthesis as a multi-objective problem where, for each scan angle,
both the radiated power density in the scan direction and the total
reflected power are accounted for by including the finite array coupling
in the design process (i.e., no periodicity approximation is employed).
The effectiveness of the approach has been assessed with a selected
set of numerical test cases by also proving the possibility to indirectly
control the \emph{ARL} of finite arrays.

\noindent The main outcomes from the numerical validation, which has
been carried out with a commercial full-wave software \cite{HFSS 2021}
to have a faithful modelling of all the electromagnetic interactions,
can be summarized as follows:

\begin{itemize}
\item \noindent the implemented design concept outperforms the standard
linear phase shifting technique \cite{Mailloux 2005} in yielding
wide-angle scanning arrays regardless of the frequency, the radiating
elements, and the array arrangement (i.e., linear or planar);
\item the \emph{PO}-based synthesis method allows one to enhance the \emph{FoV}
of finite antenna arrays without modifying the antenna architecture,
thus it can be applied to upgrade existing arrays;
\item the multi-objective design process provides the user with a multiplicity
of trade-off Pareto-front solutions, thus the possibility of choosing
an optimal one depending on an application-driven criterion based
on either functional or non-functional (e.g., costs, maintenance,
etc...) constraints.
\end{itemize}
\noindent Future works, beyond the scope of this manuscript, will
be aimed at extending the proposed methodology to conformal structures.

\section*{\noindent Acknowledgements}

\noindent This work benefited from the networking activities carried
out within the project DICAM-EXC (Departments of Excellence 2023-2027,
grant L232/2016) funded by the Italian Ministry of Education, Universities
and Research (MUR), the Project \char`\"{}Smart ElectroMagnetic Environment
in TrentiNo - SEME@TN\char`\"{} funded by the Autonomous Province
of Trento (CUP: C63C22000720003), the Project \char`\"{}AURORA - Smart
Materials for Ubiquitous Energy Harvesting, Storage, and Delivery
in Next Generation Sustainable Environments\char`\"{} funded by the
Italian Ministry for Universities and Research within the PRIN-PNRR
2022 Program, and the following projects funded by the European Union
- NextGenerationEU within the PNRR Program: Project \char`\"{}ICSC
National Centre for HPC, Big Data and Quantum Computing (CN HPC)\char`\"{}
(CUP: E63C22000970007), Project \char`\"{}Telecommunications of the
Future (PE00000001 - program \char`\"{}RESTART\char`\"{}, Structural
Project 6GWINET)'' (CUP: D43C22003080001), Project ''INSIDE-NEXT
- Indoor Smart Illuminator for Device Energization and Next-Generation
Communications'' (CUP: E53D23000990001), and Project \char`\"{}Telecommunications
of the Future (PE00000001 - program {}``RESTART'', Focused Project
MOSS)'' (CUP: J33C22002880001). A. Massa wishes to thank E. Vico
for her never-ending inspiration, support, guidance, and help.

\newpage
\section*{FIGURE CAPTIONS}

\begin{itemize}
\item \textbf{Figure 1.} \emph{Problem geometry}. Sketch of the reference
phased array layout.
\item \textbf{Figure 2.} \emph{Slot-fed patch linear array} ($N=5$, $f=2$
{[}GHz{]}, $d=0.47\lambda$) - Sketch of (\emph{a}) the geometry of
array element and (\emph{b}) the HFSS model of the array, and plot
of the Pareto-optimal solutions when $\left(\theta_{q},\varphi_{q}\right)=\left(90,52\right)$
{[}deg{]}.
\item \textbf{Figure 3.} \emph{Slot-fed patch linear array} ($N=5$, $f=2$
{[}GHz{]}, $d=0.47\lambda$, $q=53$, $\left(\theta_{q},\varphi_{q}\right)=\left(90,52\right)$
{[}deg{]}) - Plots of (\emph{a}) $\angle w_{nq}^{+}$ and (\emph{b})
$ARL_{nq}$, $n=1,...,N$.
\item \textbf{Figure 4.} \emph{Slot-fed patch linear array} ($N=5$, $f=2$
{[}GHz{]}, $d=0.47\lambda$, $q=53$, $\left(\theta_{q},\varphi_{q}\right)=\left(90,52\right)$
{[}deg{]}, $\theta=90$ {[}deg{]}) - Plot of $\psi_{q}\left(\theta,\varphi\right)$.
\item \textbf{Figure 5.} \emph{Slot-fed patch linear array} ($N=5$, $f=2$
{[}GHz{]}, $d=0.47\lambda$, $\theta_{q}=90$ {[}deg{]}) - Plots of
(\emph{a}) $\psi_{q}\left(\theta_{q},\varphi_{q}\right)$ and (\emph{b})
$\zeta_{q}$ versus $\varphi_{q}$ ($q=1,...,Q$; $Q=91$).
\item \textbf{Figure 6.} \emph{Slot-fed patch linear array} ($N=16$, $f=2$
{[}GHz{]}, $d=0.47\lambda$, $\theta_{q}=90$ {[}deg{]}) - Sketch
of (\emph{a}) the HFSS model of the array and plots of (\emph{b})
$\psi_{q}\left(\theta_{q},\varphi_{q}\right)$, and (\emph{c}) $\zeta_{q}$
versus $\varphi_{q}$ ($q=1,...,Q$; $Q=91$).
\item \textbf{Figure 7.} \emph{Slot-fed patch planar array} ($N=4\times4$,
$f=2$ {[}GHz{]}, $d_{x}=d_{y}=0.47\lambda$) - Sketch of the HFSS
model of the array.
\item \textbf{Figure 8.} \emph{Slot-fed patch planar array} ($N=4\times4$,
$f=2$ {[}GHz{]}, $d_{x}=d_{y}=0.47\lambda$) - Plots of (\emph{a})(\emph{b})
$\psi_{q}\left(\theta_{q},\varphi_{q}\right)$, (\emph{c})(\emph{d})
$\zeta_{q}$, and (\emph{e}) $\Delta\zeta_{q}^{STD}$ ($q=1,...,Q$;
$Q=361$).
\item \textbf{Figure 9.} \emph{Slot-fed patch planar array} ($N=4\times4$,
$f=2$ {[}GHz{]}, $d_{x}=d_{y}=0.47\lambda$) - Plot of $\Omega^{FoV}$
(in green) of (\emph{a}) the \emph{STD} and (\emph{b}) the \emph{PO}
arrays.
\item \textbf{Figure 10.} \emph{Pin-fed E-patch planar array} ($N=8\times8$,
$f=5.2$ {[}GHz{]}, $d_{x}=0.42\lambda$, $d_{y}=0.43\lambda$) \emph{-}
Sketch of the HFSS model of the array.
\item \textbf{Figure 11.} \emph{Pin-fed E-patch planar array} ($N=8\times8$,
$f=5.2$ {[}GHz{]}, $d_{x}=0.42\lambda$, $d_{y}=0.43\lambda$, $\varphi_{q}=0$
{[}deg{]}) \emph{-} Plots of (\emph{a}) $\psi_{q}\left(\theta_{q},\varphi_{q}\right)$
and (\emph{b}) $\zeta_{q}$ versus $\theta_{q}$ ($q=1,...,Q$; $Q=37$).
\item \textbf{Figure 12.} \emph{Pin-fed E-patch planar array} ($N=8\times8$,
$f=5.2$ {[}GHz{]}, $d_{x}=0.42\lambda$, $d_{y}=0.43\lambda$) -
Plot of $\Omega^{FoV}$ (in green) of (\emph{a}) the \emph{STD} and
(\emph{b}) the \emph{PO} arrays.
\item \textbf{Figure 13.} \emph{Pin-fed E-patch planar array} ($N=8\times8$,
$f=5.2$ {[}GHz{]}, $d_{x}=0.42\lambda$, $d_{y}=0.43\lambda$, $q=4$,
$\left(\theta_{q},\varphi_{q}\right)=\left(-75,0\right)$ {[}deg{]})
- Plots of $ARL_{nq}$, $n=1,...,N$ of (\emph{a}) \emph{PO}, (\emph{b})
\emph{SoA} \cite{Cheng.2017}\emph{,} and (\emph{c}) the \emph{STD}
arrays.
\end{itemize}

\section*{TABLE CAPTIONS}

\begin{itemize}
\item \textbf{Table I.} \emph{Pin-fed E-patch planar array} ($N=8\times8$,
$f=5.2$ {[}GHz{]}, $d_{x}=0.42\lambda$, $d_{y}=0.43\lambda$, $\varphi_{q}=0$
{[}deg{]}) - Performance indexes.
\item \textbf{Table II.} \emph{Pin-fed E-patch planar array} ($N=8\times8$,
$f=5.2$ {[}GHz{]}, $d_{x}=0.42\lambda$, $d_{y}=0.43\lambda$, $\varphi_{q}=0$
{[}deg{]}) - Performance indexes in \cite{Cheng.2017}.
\end{itemize}
\newpage
\begin{center}~\vfill\end{center}

\begin{center}\includegraphics[%
  width=0.90\columnwidth]{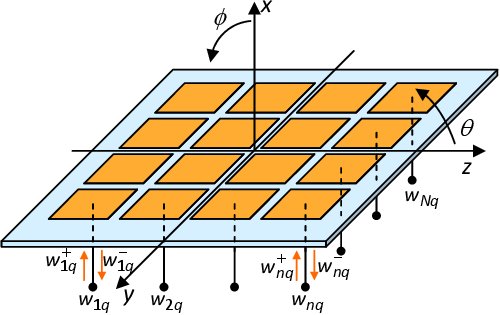}\end{center}

\begin{center}~\vfill\end{center}

\begin{center}\textbf{Fig. 1 - P. Rosatti} \textbf{\emph{et al.}}\textbf{,}
\textbf{\emph{{}``}}Synthesis of Wide-Angle Scanning Arrays ...''\end{center}

\newpage
\begin{center}~\end{center}

\begin{center}\begin{tabular}{c}
\includegraphics[%
  width=0.40\linewidth]{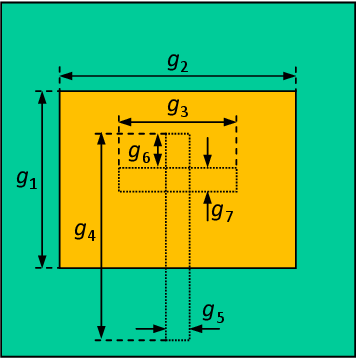}\tabularnewline
(\emph{a})\tabularnewline
\includegraphics[%
  width=0.75\columnwidth]{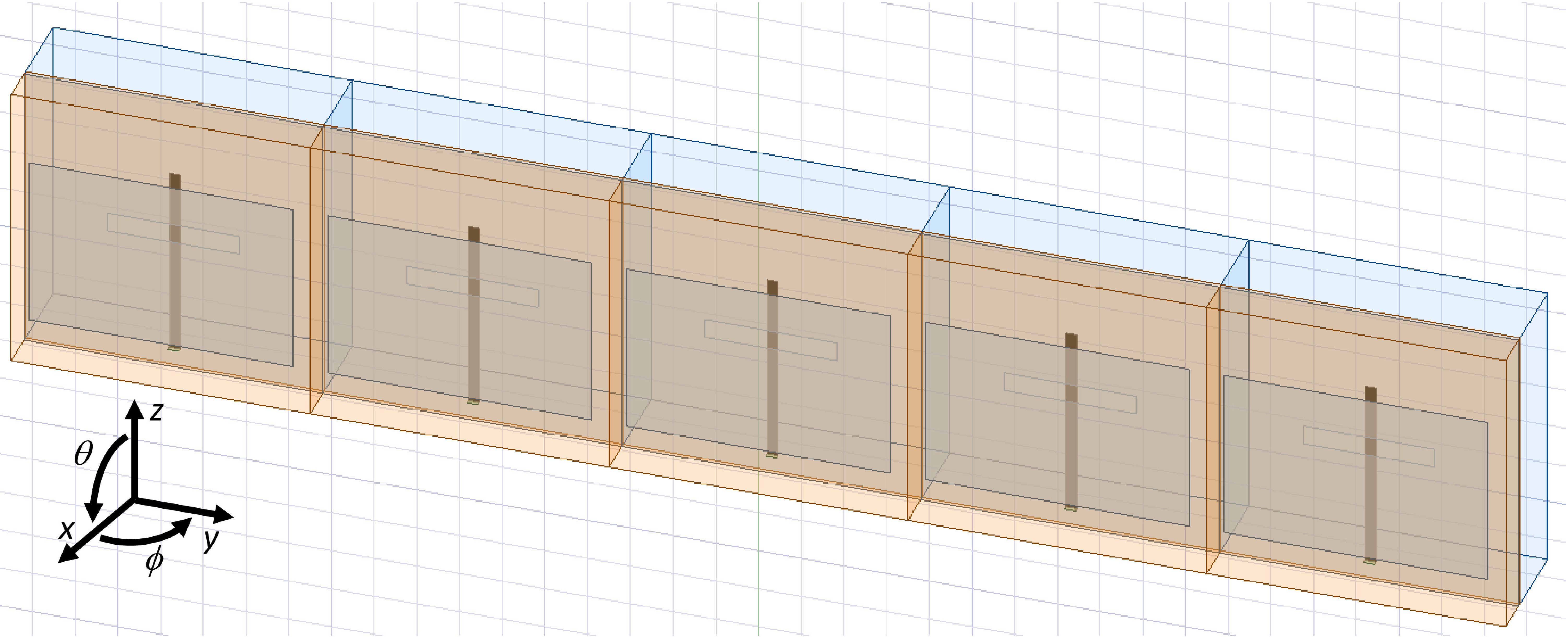}\tabularnewline
(\emph{b})\tabularnewline
\includegraphics[%
  width=0.75\columnwidth]{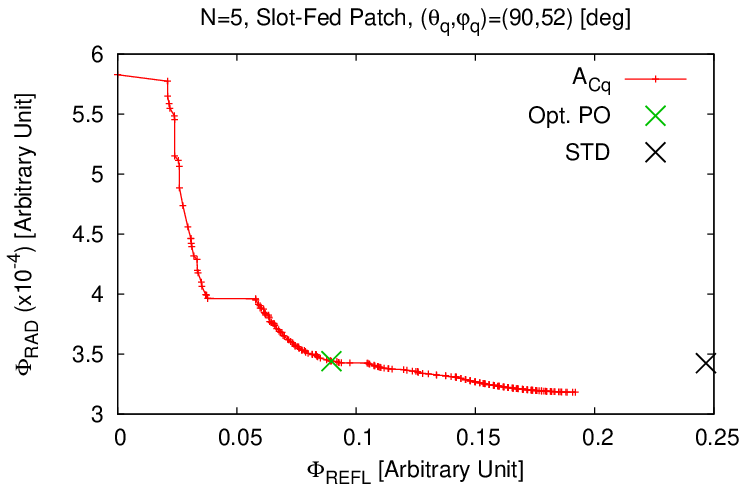}\tabularnewline
(\emph{c})\tabularnewline
\end{tabular}\end{center}

\begin{center}\textbf{Fig. 2 - P. Rosatti} \textbf{\emph{et al.}}\textbf{,}
\textbf{\emph{{}``}}Synthesis of Wide-Angle Scanning Arrays ...''\end{center}

\newpage
\begin{center}~\end{center}

\begin{center}\begin{tabular}{c}
\includegraphics[%
  width=0.80\linewidth]{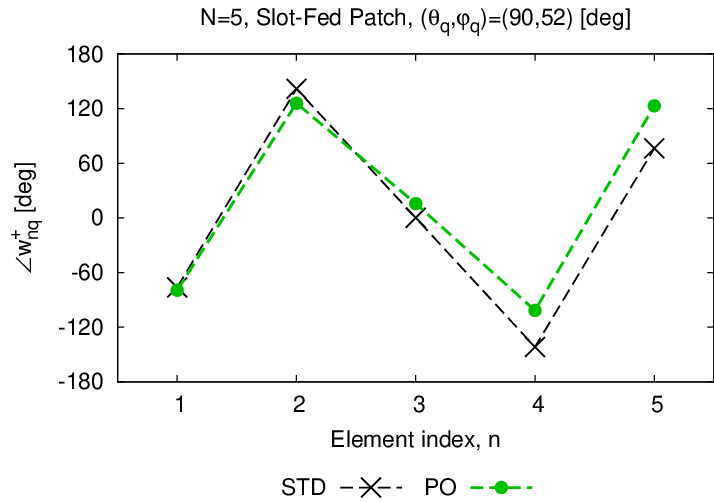}\tabularnewline
(\emph{a})\tabularnewline
\includegraphics[%
  width=0.80\linewidth]{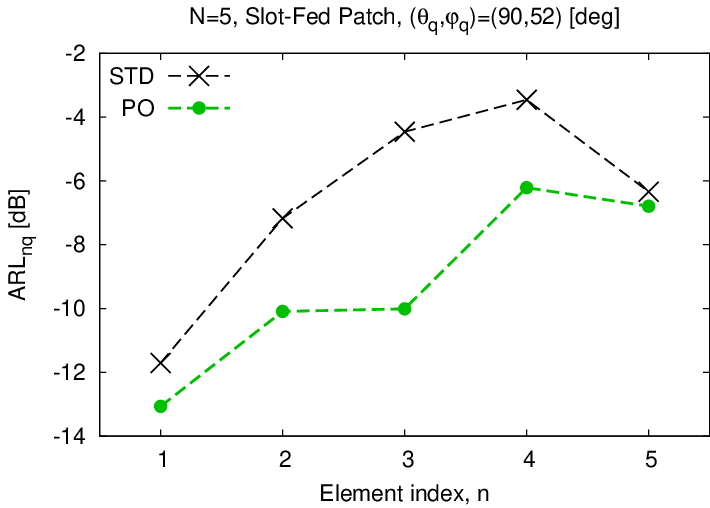}\tabularnewline
(\emph{b})\tabularnewline
\end{tabular}\end{center}

\begin{center}\vfill\end{center}

\begin{center}\textbf{Fig. 3 - P. Rosatti} \textbf{\emph{et al.}}\textbf{,}
\textbf{\emph{{}``}}Synthesis of Wide-Angle Scanning Arrays ...''\end{center}

\newpage
\begin{center}~\vfill\end{center}

\begin{center}\includegraphics[%
  width=0.90\linewidth]{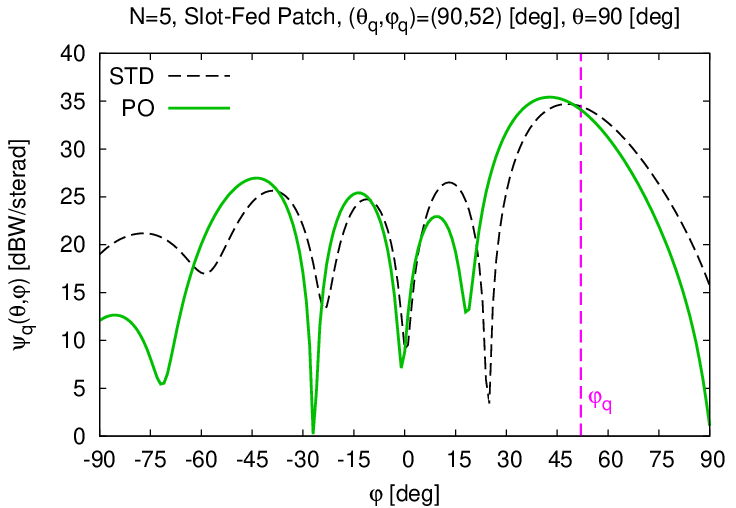}\end{center}

\begin{center}\vfill\end{center}

\begin{center}\textbf{Fig. 4 - P. Rosatti} \textbf{\emph{et al.}}\textbf{,}
\textbf{\emph{{}``}}Synthesis of Wide-Angle Scanning Arrays ...''\end{center}

\newpage
\begin{center}~\end{center}

\begin{center}\begin{tabular}{c}
\includegraphics[%
  width=0.90\columnwidth]{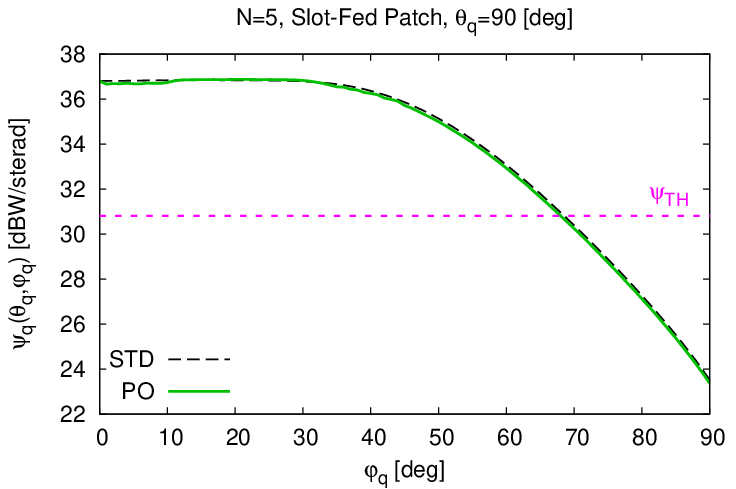}\tabularnewline
(\emph{a})\tabularnewline
\includegraphics[%
  width=0.90\columnwidth]{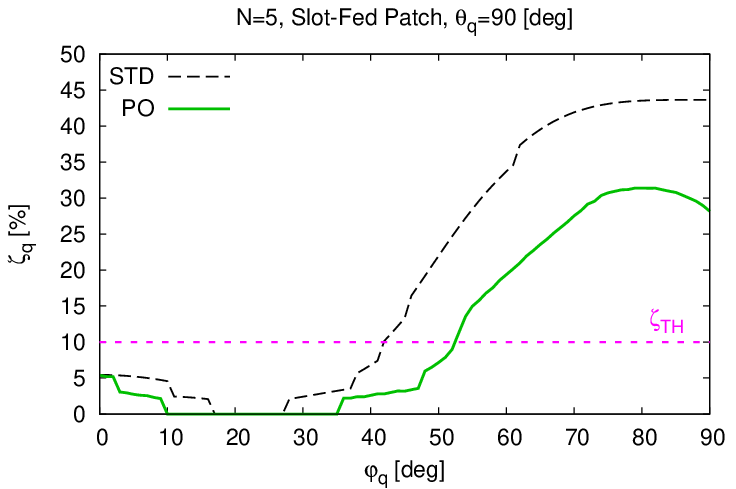}\tabularnewline
(\emph{b})\tabularnewline
\end{tabular}\end{center}

\begin{center}\vfill\end{center}

\begin{center}\textbf{Fig. 5 - P. Rosatti} \textbf{\emph{et al.}}\textbf{,}
\textbf{\emph{{}``}}Synthesis of Wide-Angle Scanning Arrays ...''\end{center}

\newpage
\begin{center}\begin{tabular}{c}
\includegraphics[%
  width=0.90\columnwidth]{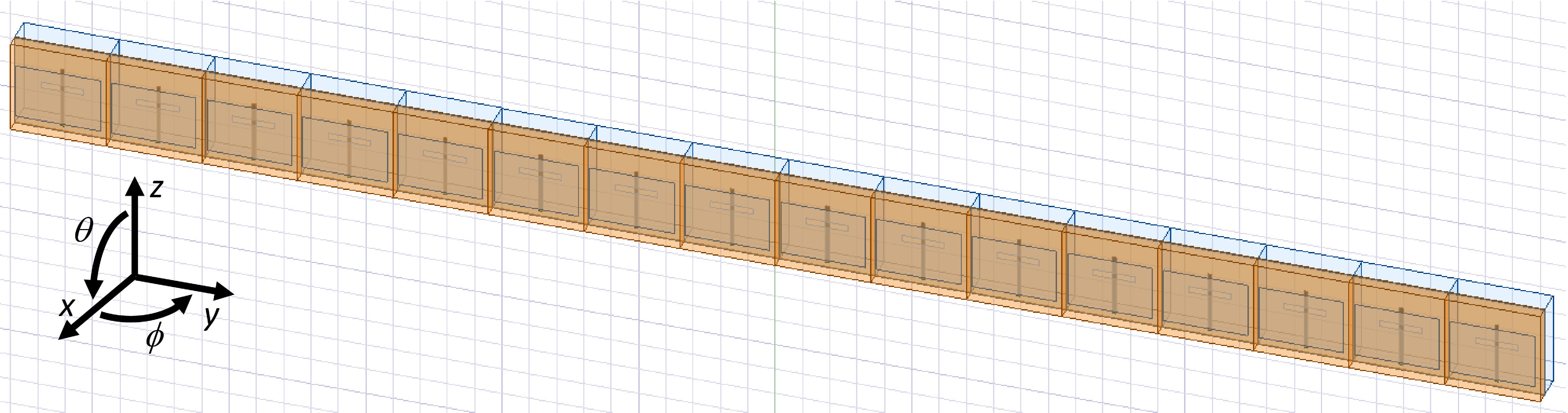}\tabularnewline
(\emph{a})\tabularnewline
\includegraphics[%
  width=0.75\linewidth]{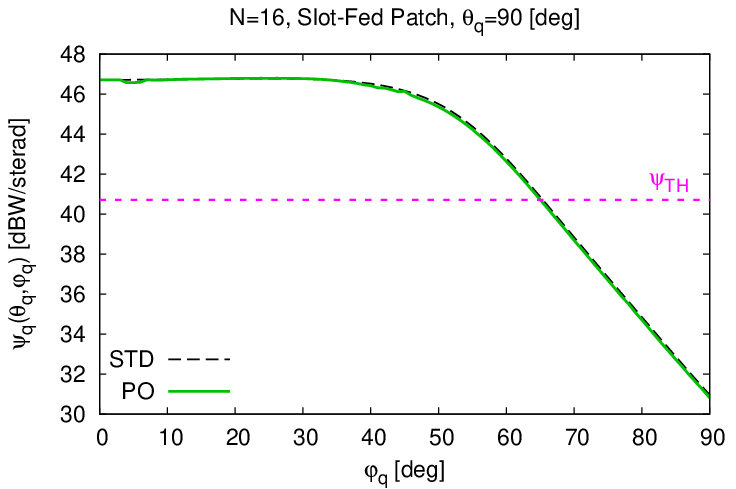}\tabularnewline
(\emph{b})\tabularnewline
\includegraphics[%
  width=0.75\linewidth]{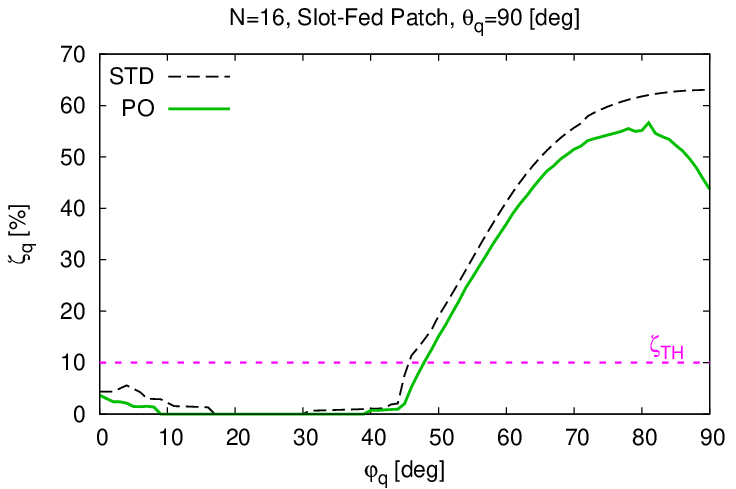}\tabularnewline
(\emph{c}) \tabularnewline
\end{tabular}\end{center}

\begin{center}\vfill\end{center}

\begin{center}\textbf{Fig. 6 - P. Rosatti} \textbf{\emph{et al.}}\textbf{,}
\textbf{\emph{{}``}}Synthesis of Wide-Angle Scanning Arrays ...''\end{center}

\newpage
\begin{center}~\vfill\end{center}

\begin{center}\includegraphics[%
  width=0.95\linewidth]{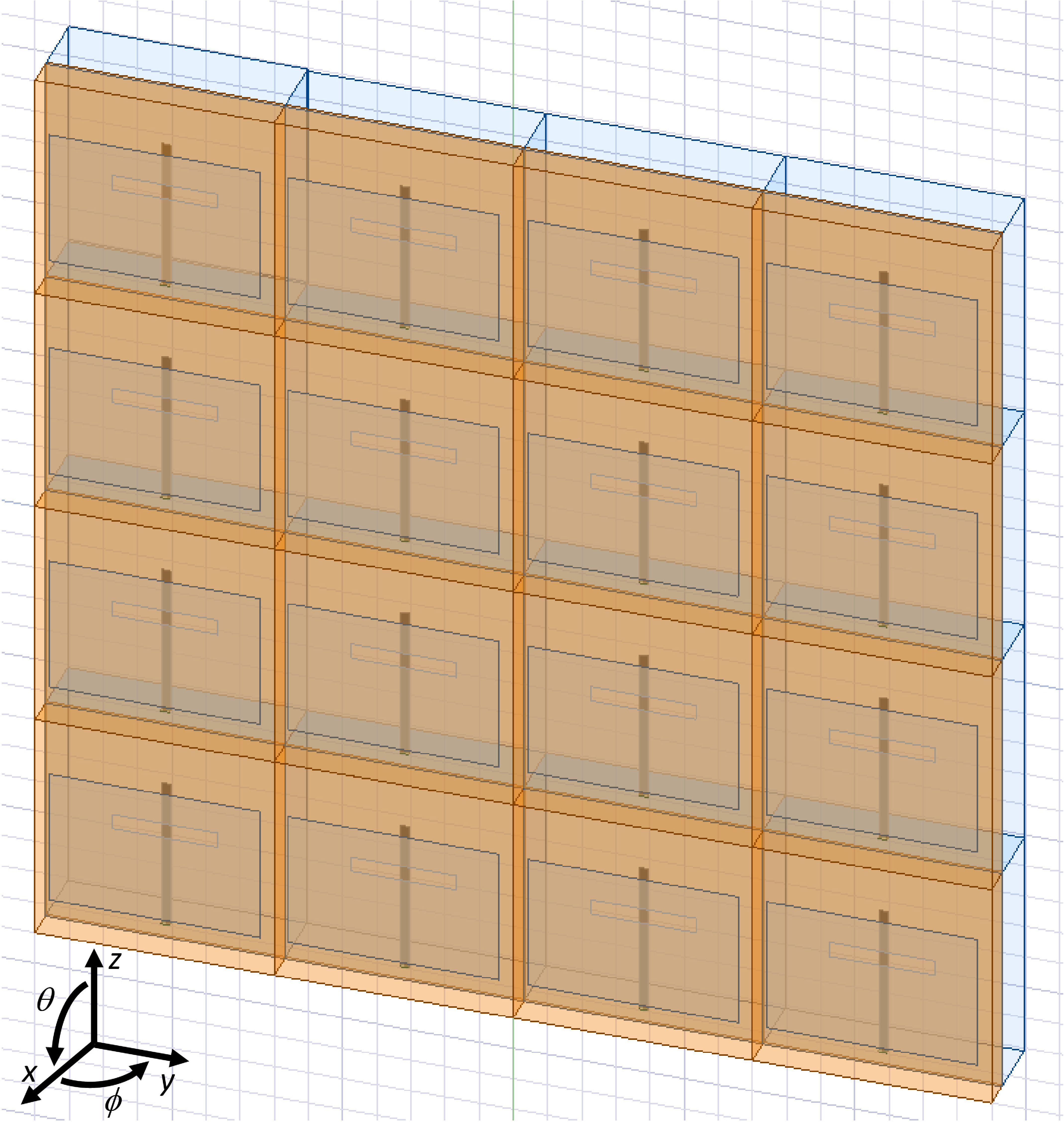}\end{center}

\begin{center}\vfill\end{center}

\begin{center}\textbf{Fig. 7 - P. Rosatti} \textbf{\emph{et al.}}\textbf{,}
\textbf{\emph{{}``}}Synthesis of Wide-Angle Scanning Arrays ...''\end{center}

\newpage
\begin{center}~\end{center}

\begin{center}\begin{tabular}{cc}
\includegraphics[%
  width=0.45\linewidth]{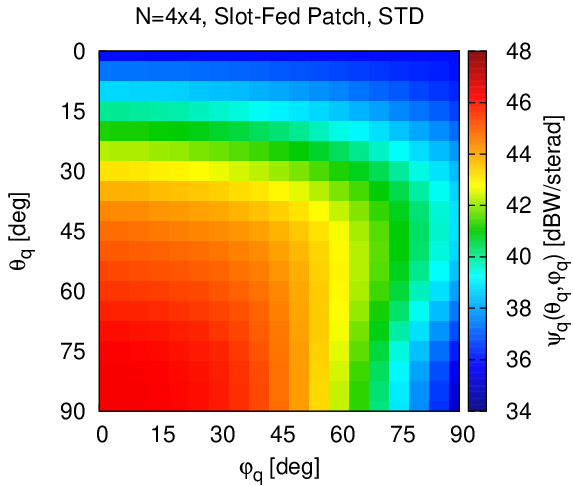}&
\includegraphics[%
  width=0.45\linewidth]{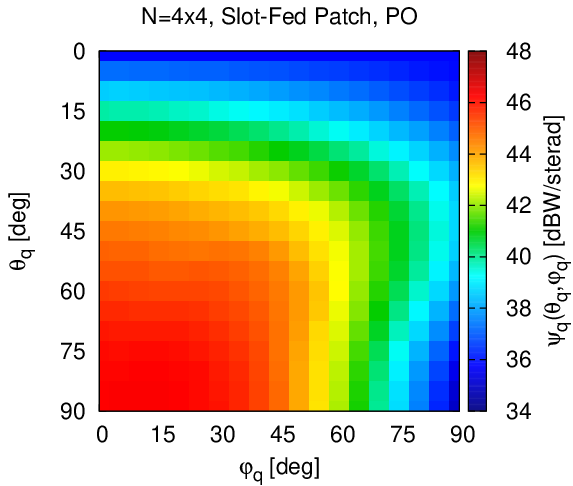}\tabularnewline
(\emph{a})&
(\emph{b})\tabularnewline
\includegraphics[%
  width=0.45\linewidth]{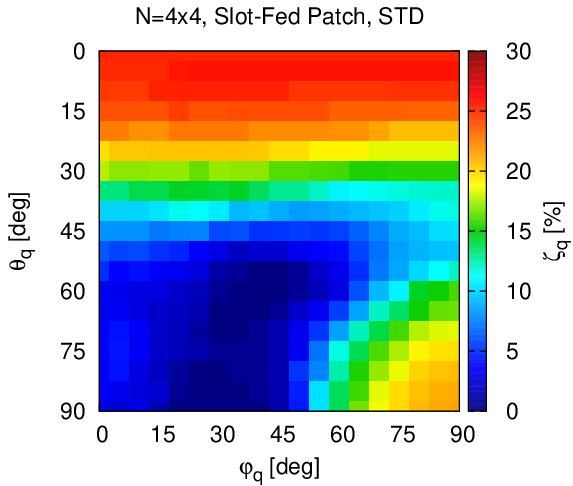}&
\includegraphics[%
  width=0.45\linewidth]{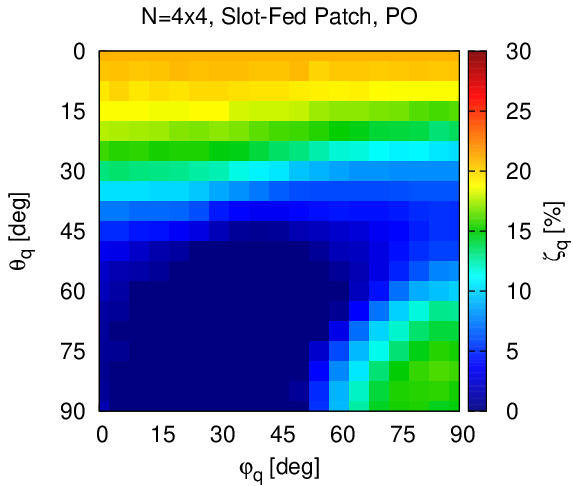}\tabularnewline
(\emph{c})&
(\emph{d})\tabularnewline
\multicolumn{2}{c}{\includegraphics[%
  width=0.45\linewidth]{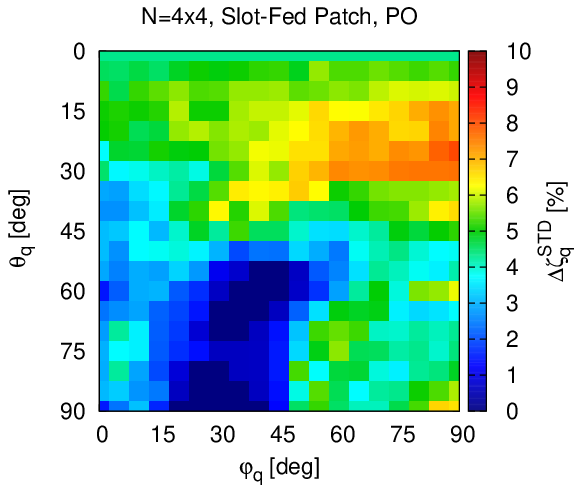}}\tabularnewline
\multicolumn{2}{c}{(\emph{e})}\tabularnewline
\end{tabular}\end{center}

\begin{center}\vfill\end{center}

\begin{center}\textbf{Fig. 8 - P. Rosatti} \textbf{\emph{et al.}}\textbf{,}
\textbf{\emph{{}``}}Synthesis of Wide-Angle Scanning Arrays ...''\end{center}

\newpage
\begin{center}\begin{tabular}{c}
\includegraphics[%
  width=0.65\linewidth]{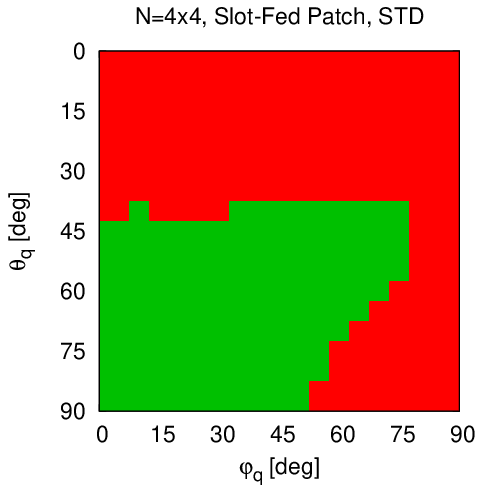}\tabularnewline
(\emph{a})\tabularnewline
\includegraphics[%
  width=0.65\linewidth]{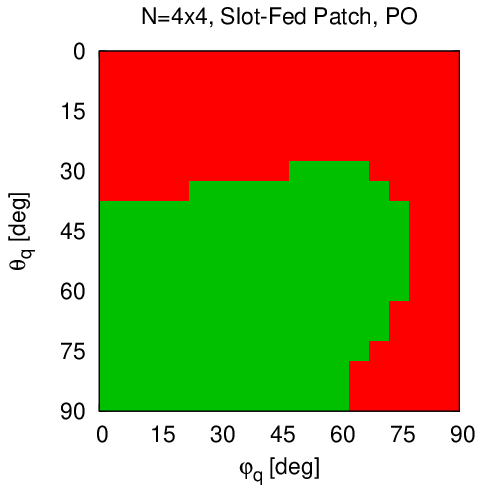}\tabularnewline
(\emph{b})\tabularnewline
\end{tabular}\end{center}

\begin{center}\textbf{Fig. 9 - P. Rosatti} \textbf{\emph{et al.}}\textbf{,}
\textbf{\emph{{}``}}Synthesis of Wide-Angle Scanning Arrays ...''\end{center}

\newpage
\begin{center}~\vfill\end{center}

\begin{center}\includegraphics[%
  width=0.90\linewidth]{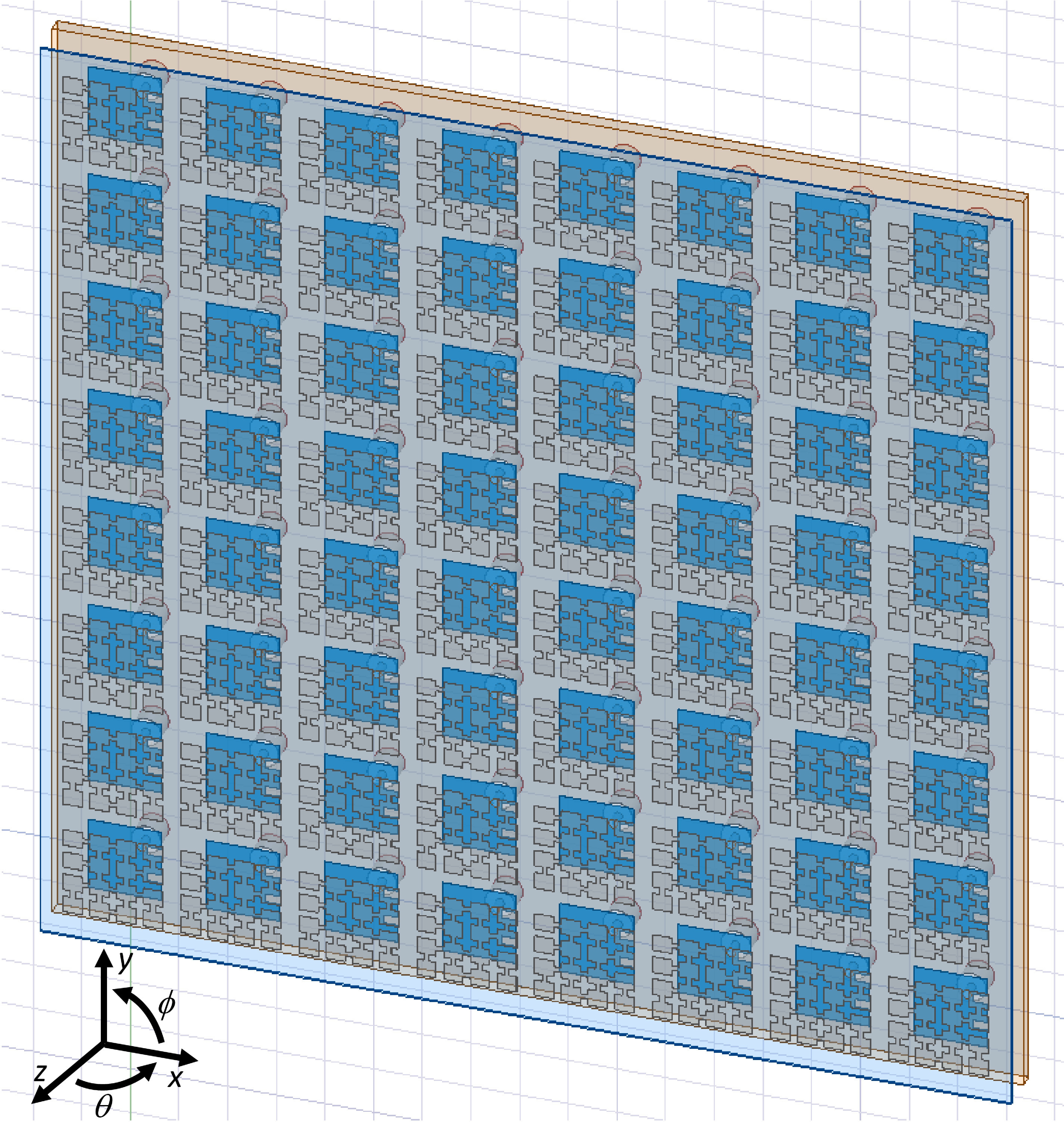}\end{center}

\begin{center}\vfill\end{center}

\begin{center}\textbf{Fig. 10 - P. Rosatti} \textbf{\emph{et al.}}\textbf{,}
\textbf{\emph{{}``}}Synthesis of Wide-Angle Scanning Arrays ...''\end{center}

\newpage
\begin{center}~\end{center}

\begin{center}\begin{tabular}{c}
\includegraphics[%
  width=0.90\linewidth]{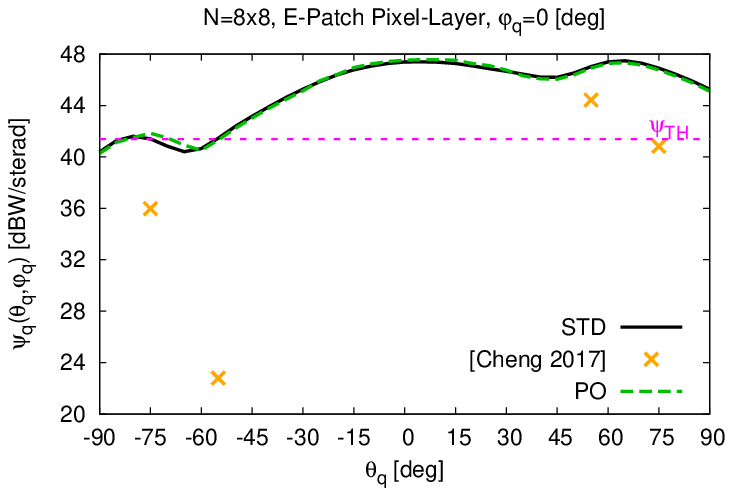}\tabularnewline
(\emph{a})\tabularnewline
\includegraphics[%
  width=0.90\linewidth]{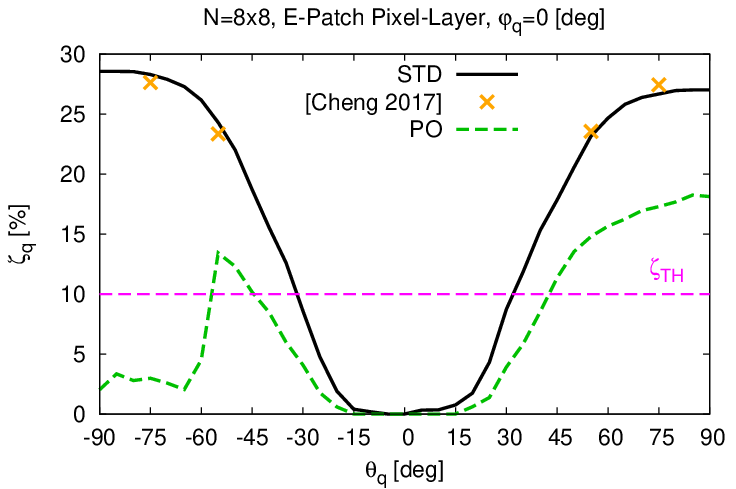}\tabularnewline
(\emph{b})\tabularnewline
\end{tabular}\end{center}

\begin{center}\vfill\end{center}

\begin{center}\textbf{Fig. 11 - P. Rosatti} \textbf{\emph{et al.}}\textbf{,}
\textbf{\emph{{}``}}Synthesis of Wide-Angle Scanning Arrays ...''\end{center}

\newpage
\begin{center}~\end{center}

\begin{center}\begin{tabular}{c}
\includegraphics[%
  width=0.90\linewidth]{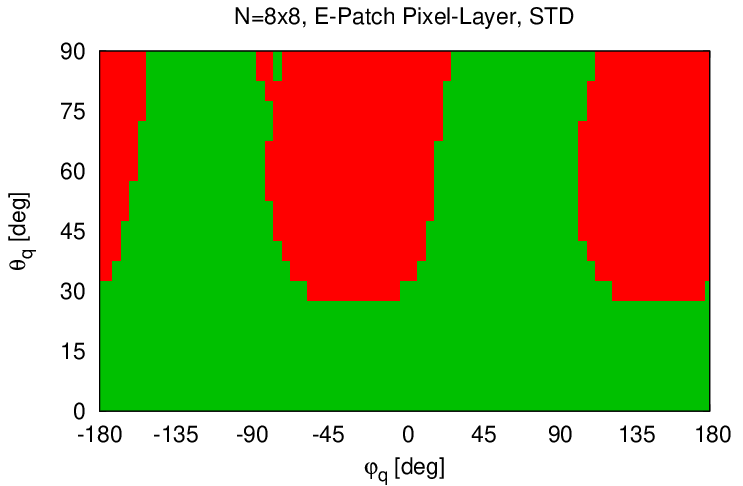}\tabularnewline
(\emph{a})\tabularnewline
\includegraphics[%
  width=0.90\linewidth]{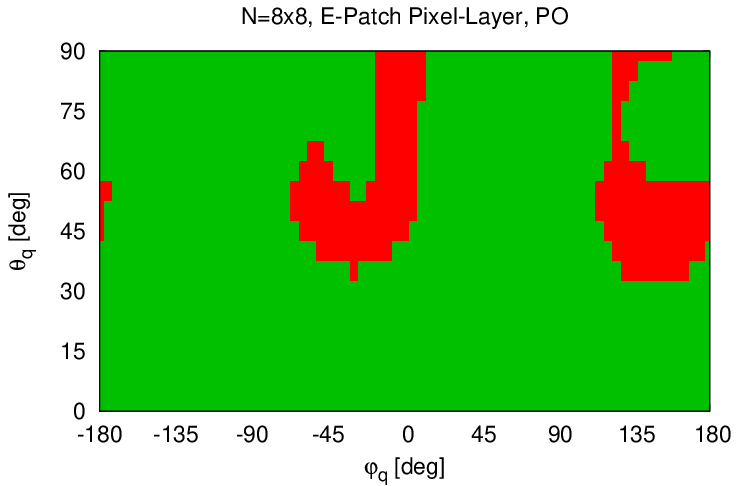}\tabularnewline
(\emph{b})\tabularnewline
\end{tabular}\end{center}

\begin{center}\vfill\end{center}

\begin{center}\textbf{Fig. 12 - P. Rosatti} \textbf{\emph{et al.}}\textbf{,}
\textbf{\emph{{}``}}Synthesis of Wide-Angle Scanning Arrays ...''\end{center}

\newpage
\begin{center}~\vfill\end{center}

\begin{center}\begin{tabular}{c}
\includegraphics[%
  width=0.48\linewidth]{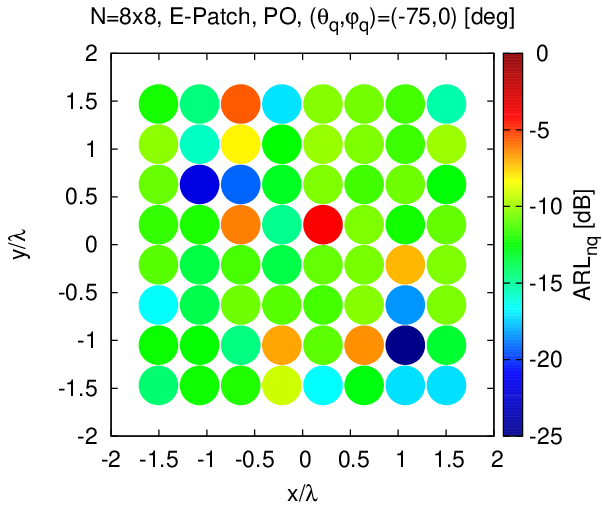}\tabularnewline
(\emph{a})\tabularnewline
\includegraphics[%
  width=0.48\linewidth]{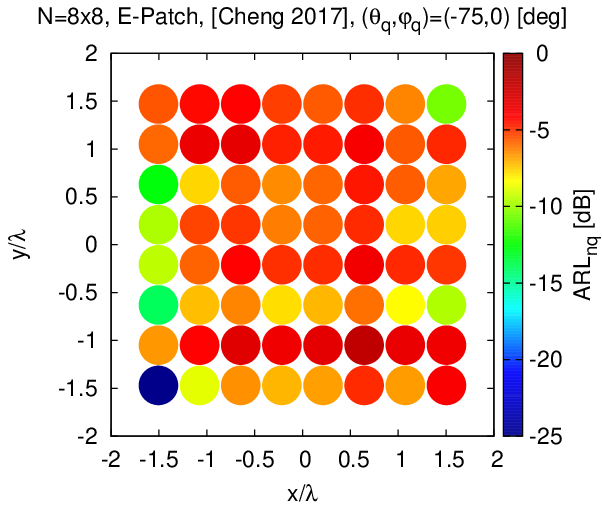}\tabularnewline
(\emph{b})\tabularnewline
\includegraphics[%
  width=0.48\linewidth,
  keepaspectratio]{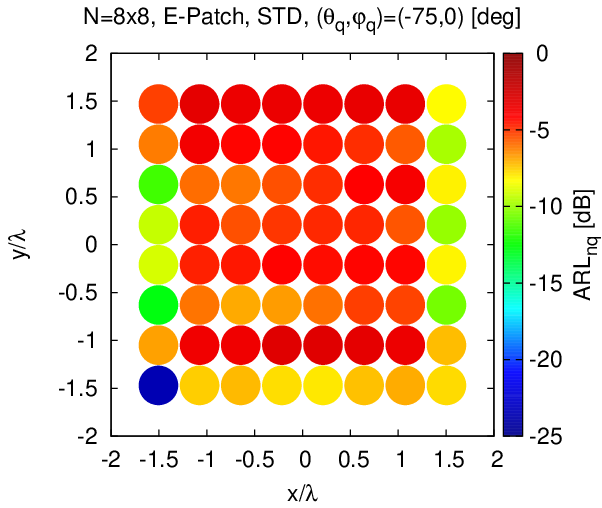}\tabularnewline
(\emph{c})\tabularnewline
\end{tabular}\end{center}

\begin{center}\textbf{Fig. 13 - P. Rosatti} \textbf{\emph{et al.}}\textbf{,}
\textbf{\emph{{}``}}Synthesis of Wide-Angle Scanning Arrays ...''\end{center}

\newpage
\begin{center}~\vfill\end{center}

\begin{center}\begin{tabular}{|c|c||c|c|c||c|c|c|}
\hline 
\textbf{$q$}&
$\left(\theta_{q},\varphi_{q}\right)$&
$\psi_{q}^{PO}\left(\theta_{q},\varphi_{q}\right)$&
$\psi_{q}^{[Cheng\,2017]}\left(\theta_{q},\varphi_{q}\right)$&
$\psi_{q}^{STD}\left(\theta_{q},\varphi_{q}\right)$&
$\zeta_{q}^{PO}$&
$\zeta_{q}^{[Cheng\,2017]}$&
$\zeta_{q}^{STD}$\tabularnewline
\hline
\hline 
$4$&
$\left(-75,0\right)$&
$41.83$&
$35.97$&
$41.40$&
$2.98$&
$27.62$&
$28.30$\tabularnewline
\hline 
$8$&
$\left(-55,0\right)$&
$41.32$&
$22.79$&
$41.45$&
$13.42$&
$23.35$&
$24.30$\tabularnewline
\hline 
$30$&
$\left(55,0\right)$&
$46.91$&
$44.42$&
$47.02$&
$14.80$&
$23.35$&
$23.16$\tabularnewline
\hline 
$34$&
$\left(75,0\right)$&
$46.76$&
$40.83$&
$46.91$&
$17.28$&
$27.43$&
$26.68$\tabularnewline
\hline
\end{tabular}\end{center}

\begin{center}~\vfill\end{center}

\begin{center}\textbf{Tab. I - P. Rosatti} \textbf{\emph{et al.}}\textbf{,}
\textbf{\emph{{}``}}Synthesis of Wide-Angle Scanning Arrays ...''\end{center}

\newpage
\begin{center}~\vfill\end{center}

\begin{center}\begin{sideways}
\begin{tabular}{|c|c||c|c|c||c|c|c||c|c|c|}
\hline 
\textbf{$q$}&
$\left(\theta_{q},\varphi_{q}\right)$&
$SLL_{q}^{PO}$&
$SLL_{q}^{[Cheng\,2017]}$&
$SLL_{q}^{STD}$&
$G_{q}^{PO}$&
$G_{q}^{[Cheng\,2017]}$&
$G_{q}^{STD}$&
$\Delta\Theta_{q}^{PO}$&
$\Delta\Theta_{q}^{[Cheng\,2017]}$&
$\Delta\Theta_{q}^{STD}$\tabularnewline
\hline
\hline 
$4$&
$\left(-75,0\right)$&
$-2.12$&
$-5.78$&
$-5.38$&
$6.57$&
$6.56$&
$7.38$&
$3.55$&
$4.03$&
$3.75$\tabularnewline
\hline 
$8$&
$\left(-55,0\right)$&
$-3.82$&
$-5.68$&
$-5.44$&
$6.54$&
$6.04$&
$7.04$&
$3.71$&
$1.86$&
$2.26$\tabularnewline
\hline 
$30$&
$\left(55,0\right)$&
$-8.83$&
$-10.06$&
$-9.85$&
$12.16$&
$11.78$&
$12.47$&
$2.77$&
$2.17$&
$2.22$\tabularnewline
\hline 
$34$&
$\left(75,0\right)$&
$-9.06$&
$-9.29$&
$-9.61$&
$12.22$&
$11.79$&
$12.62$&
$5.06$&
$0.94$&
$4.94$\tabularnewline
\hline
\end{tabular}
\end{sideways}\end{center}

\begin{center}~\vfill\end{center}

\begin{center}\textbf{Tab. II - P. Rosatti} \textbf{\emph{et al.}}\textbf{,}
\textbf{\emph{{}``}}Synthesis of Wide-Angle Scanning Arrays ...''\end{center}
\end{document}